\documentclass[aps,pra,twocolumn,groupedaddress]{revtex4-2}

\usepackage{amsmath}
\usepackage{bm}
\usepackage{graphicx}
\usepackage{hyperref}
\usepackage{times}
\hypersetup{colorlinks=true,citecolor=magenta,linkcolor=magenta,urlcolor=magenta}

\begin{document}


\title{Performance evaluation of invariant-based inverse engineering by quantum speed limit}


\author{Takuya Hatomura}
\email[]{takuya.hatomura.ub@hco.ntt.co.jp}
\affiliation{NTT Basic Research Laboratories \& NTT Research Center for Theoretical Quantum Physics, NTT Corporation, Kanagawa 243-0198, Japan}


\date{\today}

\begin{abstract}
Quantum speed limits for two time-evolved states are introduced and applied to overlap between true dynamics and approximate dynamics. 
In particular, we point out that the present idea is suitable for invariant-based inverse engineering, i.e., the worst case performance of invariant-based inverse engineering can be evaluated by using a designed time-evolved state. 
As demonstrations, we apply the present method to stimulated Raman adiabatic passage and quantum annealing, and then we find that the present idea brings valuable insight to control scheduling. 
\end{abstract}


\maketitle


\section{Introduction}
Quantum speed limits (QSLs) have been developed to discuss how fast given processes can be completed~\cite{Mandelstam1945,Margolus1998,Uhlmann1992,Deffner2017}. 
By using QSLs, we can obtain trade-off relations among distance, time duration, and energy cost. 
For example, we can discuss the amount of energy cost for transferring a given initial state to a target state within required time, or the minimum time for that within given energy cost. 
Generally speaking, realization of long distance transfer within short time requires large energy cost.

Shortcuts to adiabaticity (STA) were introduced as alternative routes offering significant speedup to a final destination of adiabatic control~\cite{Demirplak2003,Berry2009,Chen2010,Guery-Odelin2019}. 
The theory of STA itself has no limitation on speedup, i.e., it in principle enables us to mimic adiabatic time evolution within arbitrary time. 
Recently, QSLs were applied to STA for revealing physical restriction of speedup~\cite{Santos2015,Campbell2017}. 
As expected, speedup via STA is restricted by energy cost, but interestingly it was found that this energy cost is given by geometric quantities, i.e., it is given by line integral in parameter space and does not depend on time schedule on its path.

More recently, it was found that QSLs can be used for performance evaluation of quantum control~\cite{Suzuki2020,Hatomura2021,Funo2021}. 
By using the theory of QSLs, lower bounds for overlap between a time-evolved state and an adiabatic state were obtained in Ref.~\cite{Suzuki2020}. 
These bound can be used for performance evaluation of adiabatic control in realistic time scale, i.e., we can evaluate how adiabatic dynamics is. 
Moreover, a lower bound for overlap between an assisted time-evolved state, which is generated via approximate STA, and an adiabatic state was also obtained in Ref.~\cite{Hatomura2021}. 
This bound can be used for performance evaluation of approximate counterdiabatic driving, i.e., we can estimate deviations from exact counterdiabatic driving, which is one of the methods of STA~\cite{Demirplak2003,Berry2009} and offers perfect population transfer. 
A bound for evaluating the effects of environment in counterdiabatic driving was also obtained~\cite{Funo2021}.

In this paper, we discuss application of QSLs to invariant-based inverse engineering (IBIE), which is another method of STA~\cite{Chen2010}, for performance evaluation. 
For this purpose, we first introduce QSLs for two time-evolved states in a similar manner to Ref.~\cite{Suzuki2020,Hatomura2021,Funo2021}. 
Then, we assume that one of the time-evolved state is true dynamics and the other is its approximate one. 
We point out that this simple idea is very suitable for IBIE and brings valuable insight to control scheduling.

\section{Theory}
\subsection{Worst case performance of approximation}
We introduce QSLs for two time-evolved states with an identical initial state, $|\Psi_1(t)\rangle=\hat{U}_1(t)|\Psi_0\rangle$ and $|\Psi_2(t)\rangle=\hat{U}_2(t)|\Psi_0\rangle$, generated by time-dependent Hamiltonians $\hat{\mathcal{H}}_1(t)$ and $\hat{\mathcal{H}}_2(t)$, where $\hat{U}_1(t)$ and $\hat{U}_2(t)$ are time evolution operators. 
By introducing a time-evolved state, $|\Phi(t)\rangle=\hat{U}_2^\dag(t)\hat{U}_1(t)|\Psi_0\rangle$ or $|\Phi(t)\rangle=\hat{U}_1^\dag(t)\hat{U}_2(t)|\Psi_0\rangle$, we can apply the standard formalism of the QSLs~\cite{Deffner2017} to the Fubini-Study distance between $|\Psi_1(t)\rangle$ and $|\Psi_2(t)\rangle$, i.e., we obtain two QSLs
\begin{equation}
\begin{aligned}
\arccos|\langle\Psi_1(t)|\Psi_2(t)\rangle|\le\frac{1}{\hbar}\int_0^tdt^\prime\sigma[\delta\hat{\mathcal{H}}(t^\prime),|\Psi_i(t^\prime)\rangle],&\\
\text{for }i=1,2,&
\end{aligned}
\label{Eq.bound.twostates}
\end{equation}
where $\delta\hat{\mathcal{H}}(t)\equiv\hat{\mathcal{H}}_1(t)-\hat{\mathcal{H}}_2(t)$ and $\sigma[\delta\hat{\mathcal{H}},|\Psi\rangle]=\sqrt{\langle\Psi|\delta\hat{\mathcal{H}}^2|\Psi\rangle-\langle\Psi|\delta\hat{\mathcal{H}}|\Psi\rangle^2}$.

In the present paper, we assume that one of the Hamiltonians $\hat{\mathcal{H}}_1(t)$ is a true Hamiltonian, of which it is difficult to find the time-evolved state $|\Psi_1(t)\rangle$, and the other $\hat{\mathcal{H}}_2(t)$ is its approximate Hamiltonian, of which we can find the explicit expression of the time-evolved state $|\Psi_2(t)\rangle$. 
In this case, the right-hand side of a QSL [$i=2$ in Eq.~(\ref{Eq.bound.twostates})] can be calculated and it provides us with a lower bound for the overlap $|\langle\Psi_1(t)|\Psi_2(t)\rangle|$, i.e., it gives the worst case performance of approximation. 
Note that it is not always easy to find a time-evolved state even for an approximate Hamiltonian. 
However, we point out that this idea is very suitable for IBIE.

\subsection{Approximate invariant-based inverse engineering}
Here we apply IBIE to the Hamiltonian $\hat{\mathcal{H}}_2(t)$. 
Our assumption is that application of IBIE to the Hamiltonian $\hat{\mathcal{H}}_1(t)$ is difficult because of its complexity. 
According to the Lewis-Riesenfeld theory~\cite{Lewis1969}, there exist the Lewis-Riesenfeld (dynamical) invariants $\hat{F}(t)$ satisfying
\begin{equation}
i\hbar\frac{\partial}{\partial t}\hat{F}(t)-[\hat{\mathcal{H}}_2(t),\hat{F}(t)]=0, 
\label{Eq.inveq}
\end{equation}
and then the time-evolved state $|\Psi_2(t)\rangle$ is expressed as
\begin{equation}
|\Psi_2(t)\rangle=\sum_nc_n(0)e^{i\kappa_n(t)}|\phi_n(t)\rangle,
\end{equation}
where $c_n(0)$ is initial distribution, $|\phi_n(t)\rangle$ is an eigenvector of the dynamical invariant $\hat{F}(t)$, and $\kappa_n(t)$ is the Lewis-Riesenfeld phase
\begin{equation}
\kappa_n(t)=\frac{1}{\hbar}\int_0^tdt^\prime\langle\phi_n(t^\prime)|\left(i\hbar\frac{\partial}{\partial t^\prime}-\hat{\mathcal{H}}_2(t^\prime)\right)|\phi_n(t^\prime)\rangle. 
\label{Eq.LRphase}
\end{equation}
In IBIE~\cite{Chen2010}, we first design the dynamical invariant $\hat{F}(t)$ so that the initial state $|\Psi_2(0)\rangle=\sum_nc_n(0)|\phi_n(0)\rangle$ is a given initial state and the final state $|\Psi_2(T)\rangle=\sum_nc_n(T)e^{i\kappa_n(T)}|\phi_n(T)\rangle$ is a given target state, where $T$ is the operation time. 
Then we schedule time-dependence of the Hamiltonian $\hat{\mathcal{H}}_2(t)$ by using Eqs.~(\ref{Eq.inveq}) and (\ref{Eq.LRphase}).

We then drive the system described by the true Hamiltonian $\hat{\mathcal{H}}_1(t)$ by using the time schedule designed for the Hamiltonian $\hat{\mathcal{H}}_2(t)$. 
In this sense, this method can be viewed as approximate IBIE. 
The QSL (\ref{Eq.bound.twostates}), for $i=2$, provides its worst case performance without the information of the time-evolved state $|\Psi_1(t)\rangle$. 
The key point is that we do not seek for the time-evolved state $|\Psi_2(t)\rangle$ in IBIE, but we rather design it at will, and thus it matches with the present idea.

\section{Examples}
\subsection{Stimulated Raman adiabatic passage}
Now we demonstrate our method by using a simple, but important problem, and show how it works. 
As the first example, we consider stimulated Raman adiabatic passage (STIRAP)~\cite{Gaubatz1990}. 
Here, the true Hamiltonian is given by
\begin{equation}
\hat{\mathcal{H}}_1(t)=\frac{\hbar}{2}
\begin{pmatrix}
0 & \Omega_P & 0 \\
\Omega_P & 2\Delta & \Omega_S \\
0 & \Omega_S & 0
\end{pmatrix},
\end{equation}
where $\Delta$ is time-independent detuning, $\Omega_P=\Omega_P(t)$ is the pump pulse, and $\Omega_S=\Omega_S(t)$ is the Stokes pulse. 
In STIRAP, we transfer a state from $|1\rangle={}^t(1,0,0)$ to $|3\rangle={}^t(0,0,1)$ in an adiabatic way. 
The dynamical invariant of this Hamiltonian can be expressed by using the eight Gell-Mann matrices~\cite{Gell-Mann1962}, but its explicit expression would be complicated.

Now we introduce an approximate Hamiltonian
\begin{equation}
\hat{\mathcal{H}}_2(t)=\frac{\hbar}{2}
\begin{pmatrix}
0 & \Omega_P & 0 \\
\Omega_P & 0 & \Omega_S \\
0 & \Omega_S & 0
\end{pmatrix},
\end{equation}
i.e., we assume the one-photon resonance $\Delta=0$. 
For this approximate Hamiltonian, the dynamical invariant is given by~\cite{Chen2012}
\begin{equation}
\hat{F}(t)=\frac{\hbar\Omega_0}{2}
\begin{pmatrix}
0 & \cos\gamma\sin\beta & -i\sin\gamma \\
\cos\gamma\sin\beta & 0 & \cos\gamma\cos\beta \\
i\sin\gamma & \cos\gamma\cos\beta & 0
\end{pmatrix},
\end{equation}
where $\Omega_0$ is an arbitrary constant, and from Eq.~(\ref{Eq.inveq}), $\beta=\beta(t)$ and $\gamma=\gamma(t)$ satisfy auxiliary equations
\begin{equation}
\begin{aligned}
&\dot{\gamma}=\frac{1}{2}(\Omega_P\cos\beta-\Omega_S\sin\beta), \\
&\dot{\beta}=\frac{1}{2}\tan\gamma(\Omega_S\cos\beta+\Omega_P\sin\beta). 
\end{aligned}
\end{equation}
For IBIE of STIRAP, we typically use one of the eigenstates of the dynamical invariant,
\begin{equation}
|\phi_0(t)\rangle=
\begin{pmatrix}
\cos\gamma\cos\beta \\
-i\sin\gamma \\
-\cos\gamma\sin\beta
\end{pmatrix},
\end{equation}
for which the Lewis-Riesenfeld phase is given by $\kappa_0(t)=0$, i.e., $|\Psi_2(t)\rangle=|\phi_0(t)\rangle$.

The conditions for perfect population transfer are given by $\gamma(0)=0$, $\beta(0)=0$, $\gamma(T)=0$, and $\beta(T)=\pi/2$. 
However, as pointed out in Ref.~\cite{Chen2012}, these boundary conditions result in divergence of the pump and Stokes pulses because these pulses are given by
\begin{equation}
\begin{aligned}
&\Omega_P=2(\dot{\beta}\cot\gamma\sin\beta+\dot{\gamma}\cos\beta), \\
&\Omega_S=2(\dot{\beta}\cot\gamma\cos\beta+\dot{\gamma}\sin\beta). 
\end{aligned}
\end{equation}
Therefore, in Ref.~\cite{Chen2012} the authors adopt boundary conditions $\gamma(0)=\epsilon$, $\beta(0)=0$, $\gamma(T)=\epsilon$, and $\beta(T)=\pi/2$, where $\epsilon$ is a small constant, $\epsilon\ll1$. 
In this case, the final fidelity to $|3\rangle$ becomes $(\cos\epsilon)^2\approx1-\epsilon^2$.

We adopt the protocol 1 of Ref.~\cite{Chen2012}, i.e.,
\begin{equation}
\gamma(t)=\epsilon,\quad\beta(t)=\frac{\pi t}{2T},
\end{equation}
and introduce $\Omega_\mathrm{max}=\pi/T\epsilon$, which is the maximum value of $\Omega_P$ and $\Omega_S$ for $\epsilon\ll1$. 
Then, the bound (\ref{Eq.bound.twostates}) for $i=2$ gives a lower bound for the overlap, i.e., the worst case performance of approximate IBIE,
\begin{equation}
\begin{aligned}
|\langle\Psi_1(T)|\Psi_2(T)\rangle|&\ge\cos[\Delta T\sin(2\epsilon)] \\
&\simeq\cos\frac{2\pi\Delta}{\Omega_\mathrm{max}},
\end{aligned}
\end{equation}
where the first line is the exact bound and the second line is approximate equality for $\epsilon\ll1$. 
Therefore, when $\Delta/\Omega_\mathrm{max}\ll1$, the time-evolved state $|\Psi_1(t)\rangle$ generated by the true Hamiltonian $\hat{\mathcal{H}}_1(t)$ is well-described by the designed time-evolved state $|\Psi_2(t)\rangle$, and thus we can realize high fidelity population transfer.

\subsection{Quantum annealing}
Next, we apply our method to a little bit complicated problem, and show that our method can be used for improving control schedules. 
As the second example, we consider quantum annealing~\cite{Kadowaki1998} in an infinite-range Ising model. 
Note that we set $\hbar=1$ for this example to adopt the conventional notation of quantum annealing. 
Here, the true Hamiltonian is given by
\begin{equation}
\hat{\mathcal{H}}_1(t)=A(t)\hat{\mathcal{H}}_P+B(t)\hat{\mathcal{H}}_V,
\end{equation}
where
\begin{equation}
\begin{aligned}
&\hat{\mathcal{H}}_P=-\frac{J}{2N}\sum_{i,j=1}^N\hat{Z}_i\hat{Z}_j-h\sum_{i=1}^N\hat{Z}_i, \\
&\hat{\mathcal{H}}_V=-\Gamma\sum_{i=1}\hat{X}_i,
\end{aligned}
\end{equation}
are the problem Hamiltonian and the driver Hamiltonian, and $A(t)$ and $B(t)$ are time-dependent parameters. 
Here, $J$ is coupling strength, $h$ is a longitudinal field, $\Gamma$ is a transverse field, $N$ is the number of qubits, and $\{\hat{X}_i,\hat{Y}_i,\hat{Z}_i\}_{i=1,2,\dots,N}$ are the Pauli matrices for the $N$ qubits. 
In standard quantum annealing, we consider population transfer from the ground state of the driver Hamiltonian $\hat{\mathcal{H}}_V$ to the ground state of the problem Hamiltonian $\hat{\mathcal{H}}_P$. 
Note that, owing to the permutation symmetry of the present model and assumption of the initial state, we can rewrite the present problem in terms of collective spins
\begin{equation}
\hat{S}_W=\sum_{i=1}^N\hat{W}_i/2,\quad W=X,Y,Z, 
\end{equation}
in the maximum spin subspace $\hat{\bm{S}}^2=\sum_{W=X,Y,Z}\hat{S}_W^2=(N/2)(N/2-1)$.

As the approximate Hamiltonian, we introduce the mean-field approximate Hamiltonian
\begin{equation}
\hat{\mathcal{H}}_2(t)=A(t)\hat{\mathcal{H}}_P^\mathrm{MF}+B(t)\hat{\mathcal{H}}_V,
\end{equation}
where
\begin{equation}
\hat{\mathcal{H}}_P^\mathrm{MF}=-2\left(\frac{2JM_Z}{N}+h\right)\hat{S}_Z+\frac{2JM_Z^2}{N}. 
\end{equation}
Here, $M_Z$ is the time-dependent mean-field $M_Z=M_Z(t)=\langle\Psi_2(t)|\hat{S}_Z|\Psi_2(t)\rangle$. 
For this approximate Hamiltonian, the dynamical invariant is given by~\cite{Takahashi2017}
\begin{equation}
\hat{F}(t)=h_0(\sin\beta\cos\gamma\hat{S}_X+\sin\beta\sin\gamma\hat{S}_Y+\cos\beta\hat{S}_Z), 
\end{equation}
where $h_0$ is an arbitrary constant, and from Eq.~(\ref{Eq.inveq}), $\beta=\beta(t)$ and $\gamma=\gamma(t)$ satisfy auxiliary equations
\begin{equation}
\begin{aligned}
&\dot{\beta}=-2B(t)\Gamma\sin\gamma, \\
&\dot{\gamma}=-2A(t)\left(\frac{2JM_Z}{N}+h\right)+2B(t)\Gamma\cot\beta\cos\gamma. 
\end{aligned}
\end{equation}
Here we adopt one of the eigenstates of the dynamical invariant,
\begin{equation}
\begin{aligned}
&|\phi_{N/2}(t)\rangle=\sum_{n=0}^N\tilde{c}_n(t)|-N/2+n\rangle, \\
&\tilde{c}_n(t)=\sqrt{\binom{N}{n}}\left(\cos\frac{\beta}{2}\right)^n\left(e^{i\gamma}\sin\frac{\beta}{2}\right)^{N-n},
\end{aligned}
\end{equation}
where $|m\rangle$ is the eigenstate of the collective spin operator $\hat{S}_Z$, i.e., $\hat{S}_Z|m\rangle=m|m\rangle$. 
The time-evolved state is given by $|\Psi_2(t)\rangle=e^{i\kappa_{N/2}(t)}|\phi_{N/2}(t)\rangle$. 
Note that the Lewis-Riesenfeld phase $\kappa_{N/2}(t)$ does not affect final conclusion and the mean-field is now given by
\begin{equation}
M_Z=\frac{N}{2}\cos\beta. 
\end{equation}

The conditions for standard quantum annealing are given by $\gamma(0)=0$, $\beta(0)=\pi/2$, and $\beta(T)=0$, but these boundary conditions again result in divergence of the annealing schedules because the annealing schedules are given by
\begin{equation}
\begin{aligned}
&A(t)=-\frac{\dot{\gamma}+\dot{\beta}\cot\beta\cot\gamma}{2(J\cos\beta+h)} \\
&B(t)=-\frac{\dot{\beta}}{2\Gamma\sin\gamma}. 
\end{aligned}
\end{equation}
In Ref.~\cite{Takahashi2017}, polynomial schedules are introduced for avoiding divergence of the annealing schedules, but we rather introduce small deviations as in the case of the previous example, i.e., we adopt the boundary conditions $\gamma(0)=\epsilon_\gamma$, $\beta(0)=\pi/2$, and $\beta(T)=\epsilon_\beta$, where $\epsilon_\gamma$ is an arbitrary constant and $\epsilon_\beta$ is a small constant, $\epsilon_\beta\ll1$. 
In this case, the final fidelity to the target state $|N/2\rangle$ becomes $[\cos(\epsilon_\beta/2)]^{2N}\approx1-N\epsilon_\beta^2/4$, i.e., $\epsilon_\beta$ must be much smaller than $1/\sqrt{N}$.

\begin{figure}
\includegraphics[width=8cm]{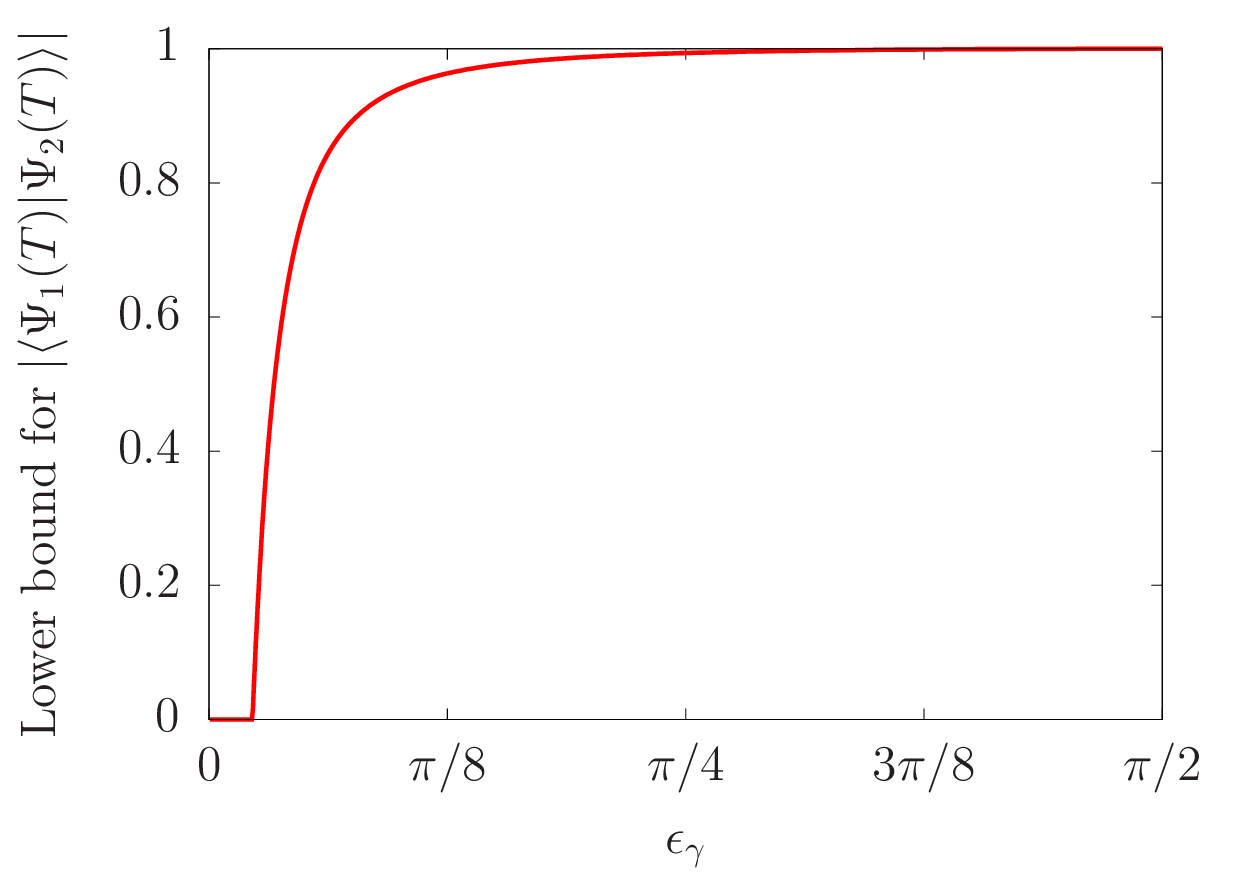}
\caption{\label{Fig.bdforqa} Lower bound for the overlap $|\langle\Psi_1(T)|\Psi_2(T)\rangle|$ against the twisting angle $\epsilon_\gamma$. The horizontal axis is the angle $\epsilon_\gamma$ and the vertical axis is the lower bound. Here, $h/J=1$, $\epsilon_\beta=0.01$, and $N=100$. }
\end{figure}

In this paper, we adopt the linear protocol
\begin{equation}
\gamma(t)=\epsilon_\gamma,\quad\beta(t)=\frac{\pi}{2}\left[1-\left(1-\frac{2\epsilon_\beta}{\pi}\right)\frac{t}{T}\right]. 
\end{equation}
Then, the bound (\ref{Eq.bound.twostates}) for $i=2$ is given by
\begin{equation}
\begin{aligned}
|\langle\Psi_1(T)|\Psi_2(T)\rangle|\ge&\cos\left[\frac{1}{2\sqrt{2}}\int_0^Tdt\frac{|\dot{\beta}|\cos\beta|\cot\epsilon_\gamma|}{\cos\beta+h/J}\right. \\
&\left.\times\sqrt{\sin^2\beta-\frac{1}{N}(\sin^2\beta-2\cos^2\beta)}\right]. 
\end{aligned}
\label{Eq.bd.qa}
\end{equation}
Although performing this integral in an analytical way is difficult unlike the previous example, we can find a way for increasing the lower bound, i.e., a way for improving the worst case performance. 
We can increase the lower bound by setting $\epsilon_\gamma\to\pi/2$, i.e., by twisting the initial state along the $z$-axis, we can improve the worst case performance. 
Here we plot how the lower bound increases against the twist in Fig.~\ref{Fig.bdforqa}. 
Surprisingly, twisting by small angle, e.g., $\epsilon_\gamma\sim\pi/8$, drastically improves the worst case performance, i.e., the time-evolved state $|\Psi_1(t)\rangle$ generated by the true Hamiltonian $\hat{\mathcal{H}}_1(t)$ is well-described by the designed time-evolved state $|\Psi_2(t)\rangle$, and thus we can realize high fidelity population transfer.

Note that for simplicity we adopt the linear protocol, and thus the present scheme is  not conventional quantum annealing. 
Indeed, $A(0)=0$, $B(0)=(\pi/2-\epsilon_\beta)/2\Gamma T\sin\epsilon_\gamma$, and $A(T)=(\pi/2-\epsilon_\beta)\cot\epsilon_\beta\cot\epsilon_\gamma/2(J\cos\epsilon_\beta+h)$, but $B(T)=(\pi/2-\epsilon_\beta)/2\Gamma T\sin\epsilon_\gamma$. 
If one would like to avoid final excitations due to the finite transverse field at the final time, appropriate protocols should be adopted so that $\dot{\beta}(T)=0$. 
Notably, we do not use the detailed schedule of $\beta$ in the derivation of the bound (\ref{Eq.bd.qa}), and thus the above conclusion holds even for such protocols.

\section{Summary}
We applied the theory of quantum speed limits to overlap between two time-evolved states with true dynamics and approximate dynamics in mind. 
We found that this idea is useful for evaluating performance of approximate invariant-based inverse engineering, where time-dependence of a true Hamiltonian is designed by using a dynamical invariant of its approximate Hamiltonian. 
Because a time-evolved state is first designed in invariant-based inverse engineering in a desired way, the bound for the overlap can easily be calculated. 
We applied this method to stimulated Raman adiabatic passage and quantum annealing, and found the conditions for high-fidelity population transfer. 
We expect that similar conditions could also be obtained for other systems using approximate invariant-based inverse engineering.

\bibliography{QSLforIBIEbib}

\end{document}